\documentclass[prc,twocolumn,superscriptaddress,superscriptfootnote,nofootinbib]{revtex4}
\usepackage{amssymb}

\usepackage{amsmath,amsfonts,amssymb,epsfig,color,MnSymbol}

\begin{document}
\title{Symplectic symmetry approach to clustering in atomic nuclei: The case of $^{24}$Mg}
\author{H. G. Ganev}
\affiliation{Joint Institute for Nuclear Research, Dubna, Russia}
\affiliation{Institute of Mechanics, Bulgarian Academy of Sciences,
Sofia, Bulgaria}

\setcounter{MaxMatrixCols}{10}

\begin{abstract}
Symplectic symmetry approach to clustering (SSAC) in atomic nuclei,
recently proposed, is modified and further developed in more detail.
It is firstly applied to the light two-cluster $^{20}$Ne + $\alpha$
system of $^{24}$Mg, the latter exhibiting well developed low-energy
$K^{\pi} = 0^{+}_{1}$, $K^{\pi} = 2^{+}_{1}$ and $K^{\pi} =
0^{-}_{1}$ rotational bands in its spectrum. A simple algebraic
Hamiltonian, consisting of dynamical symmetry, residual and vertical
mixing parts is used to describe these three lowest rotational bands
of positive and negative parity in $^{24}$Mg. A good description of
the excitation energies is obtain by considering only the $SU(3)$
cluster states restricted to the stretched many-particle Hilbert
subspace, built on the leading Pauli allowed $SU(3)$ multiplet for
the positive- and negative-parity states, respectively. The coupling
to the higher cluster-model configurations allows to describe the
known low-lying experimentally observed $B(E2)$ transition
probabilities within and between the cluster states of the three
bands under consideration without the use of an effective charge.
\end{abstract}
\maketitle

PACS number(s): {21.10.Re, 21.60.Gx, 21.60.Fw, 23.20.Lv, 23.20.-g,
27.30.+t}

\section{Introduction}

It is known that in nuclear structure physics there are three
fundamental models
\cite{RW,Ring80,Wildermuth77,nmolecules95,BM,Obertelli21,Kota20}
describing nuclear states, which are based on different physical
ideas about the structure of the nucleus. These models, in their
simplest formulation, are the shell model \cite{Heyde94}, the
cluster model \cite{Wildermuth77,nmolecules95,Clusters1} and the
collective model \cite{BM}. The shell model suggests that the atomic
nucleus is something like a small atom, the cluster model suggests
that it is like a molecule, and the collective model says that it is
like a microscopic liquid drop. Decay properties and nuclear
reactions are most naturally interpreted within the framework of
different cluster models based on the molecular picture of the
structure of atomic nucleus. The most fundamental of these three
models is, however, the nuclear shell model representing the atomic
nucleus as a proton-neutron nuclear system endowed with microscopic
antisymmetric wave functions and described within the framework of
standard multifermion quantum mechanics. The shell model, in a wider
sense, provides a general microscopic framework, in which other
models of nuclear structure can be founded and microscopically
interpreted.

The main connections between the three types of nuclear structure
models were discovered already in the 1950's. Elliott
\cite{Elliott58} has showed how quadrupole deformation and
collective rotation can be derived from the spherical shell model,
in which the collective states of the rotational band are determined
by their $SU(3)$ symmetry. Wildermuth and Kanellopoulos
\cite{Wildermuth58}, in turn, have established a connection between
the cluster and shell models based on the Hamiltonian of a
three-dimensional harmonic oscillator. In this case, the wave
function of the one model is expressed as a linear combination of
the wave functions of the another model. This connection has been
interpreted by Bayman and Bohr \cite{Bayman58} in terms of the
$SU(3)$ symmetry. As a consequence of this, the cluster states are
also selected from the model space of the shell model by their
specific $SU(3)$ symmetry. In this way, the mutual relationship
between the three fundamental models of nuclear structure has been
established in terms of the $SU(3)$ symmetry for the case of a
single shell. Recently, such a connection between the shell, cluster
and collective models has been established for the multi-shell case
using the $U_{x}(3) \otimes U_{y}(3) \supset U(3)$ dynamical
symmetry chain \cite{Cseh15,Cseh21}, which appears to be a common
intersection. All this demonstrates the crucial role played by the
$U(3)$ (or $SU(3)$) symmetry of the three-dimensional harmonic
oscillator.

There are various types of cluster models, which can be separated
into two types $-$ phenomenological and microscopic. In different
cluster models the corresponding degrees of freedom can be divided
into two categories, related to: 1) the relative motion of the
clusters; and 2) their internal structure. Especially powerful are
the algebraic models, based on the use of spectrum generating
algebras (SGA) and dynamical groups \cite{DGSGA}. Further we will
restrict our consideration only to the algebraic nuclear models.

In algebraic models all model observables, such as Hamiltonian and
transition operators, are expressed in terms of the elements of a
Lie algebra of observables. Moreover, the symmetry approach appears
as an unifying concept of different nuclear structure models by
exploiting their algebraic structures. As was mention, the most
fundamental model of nuclear structure is the nuclear shell model
(see, e.g., \cite{Heyde94}). Then, a certain nuclear structure model
becomes a submodel of the shell model if its dynamical group is
expressed as a subgroup of a dynamical group of the shell model. The
full Lie algebra of observables of the shell model is huge
(actually, infinite), which is the reason for making the shell model
(with major-shell mixing) an unsolvable problem and for seeking of
its tractable approximations. Fortunately, it has a subalgebra which
is easier to manage $-$ the Lie algebra of all one-body operators.
The corresponding dynamical group is then the group of one-body
unitary transformations. An example of a complete algebraic model
that is a submodel of the shell model is provided by the Elliott
$SU(3)$ model \cite{Elliott58} of nuclear rotations for the light
$sd$-shell nuclei. Another example is provided by the embedding of
the Bohr-Mottelson collective model \cite{BM} into the one-
\cite{stretched,Rowe96} or two-component proton-neutron
\cite{microBM,mbm-cr} microscopic shell model nuclear theory.

Symmetry, particularly the permutational symmetry, allows also to
distinguish the phenomenological nuclear structure models from the
microscopic ones. It is well known that a characteristic feature
that distinguishes between the two groups (phenomenological and
microscopic) is provided by the Pauli principle. The models are
referred to as microscopic if they fulfil the Pauli principle, which
originates from the fermion nature of atomic nucleus. For
phenomenological models the situation is opposite $-$ they do not
respect the Pauli principle, i.e. the composite fermion structure of
the nucleus is not taken into account. The important role of the
Pauli principle in the cluster models of nuclei has well been
demonstrated recently in Ref. \cite{Pauli-cluster}. Examples of
algebraic phenomenological cluster models are provided by the
nuclear vibron model (NVM)
\cite{cluster1,cluster2,cluster3,cluster4,cluster5}, the algebraic
cluster model (ACM) \cite{Bijker14,D3h-12C,Bijker19,Bijker20},
whereas the microscopic cluster models are represented by the
semimicroscopic algebraic cluster model (SACM) \cite{SACMa,SACMb},
the semimicroscopic algebraic quartet model (SAQM) \cite{SAQM}.

Recently, a microscopic algebraic cluster model based on the
symplectic symmetry has been proposed \cite{Ganev21c}. In the latter
work, considering the $^{16}$O + $\alpha \rightarrow$ $^{20}$Ne
two-cluster system, the equivalence of the new approach to the SACM
has been demonstrated. The approach of Ref.\cite{Ganev21c} has not
been, however, yet practically applied to the description of cluster
states in specific nuclear system. Thus, it is the purpose of the
present work to test the new approach, which we term the symplectic
symmetry approach to clustering (SSAC), to a real nuclear system. As
a such system, we choose the two-cluster system $^{20}$Ne + $\alpha
\rightarrow$ $^{24}$Mg. In contrast to the case of $^{16}$O +
$\alpha \rightarrow$ $^{20}$Ne considered in \cite{Ganev21c}, one of
the clusters here is characterized by a non-scalar $U_{C}(3)$
representation. In addition, $^{24}$Mg exhibits in the
experimentally observed low-lying spectrum well developed $K^{\pi} =
0^{+}_{1}$, $K^{\pi} = 2^{+}_{1}$ and $K^{\pi} = 0^{-}_{1}$
rotational bands. The nucleus $^{24}$Mg has been studied within the
framework of different algebraic nuclear structure models. For
instance, it has been studied long time ago within the nuclear
$SU(3)$ shell model \cite{Harvey68}. In Refs.\cite{Mg24a,Mg24b} the
low-lying structure of the ground and $\gamma$ rotational bands in
$^{24}$Mg has been investigated within the framework of the
microscopic $Sp(6,R)$ symplectic model (sometimes referred to as a
microscopic collective model), whereas in Ref.\cite{Mg24c} this
nucleus has been studied within the SACM considering the single
$^{12}$C $+$ $^{12}$C $\rightarrow$ $^{24}$Mg channel, including
both the low-lying and the high-energy quasimolecular resonance
states. For different cluster model approaches to $^{24}$Mg see,
e.g., the detailed list of relevant references given in
\cite{Mg24c}. In the present work we consider only the low-energy
excited states of the lowest $K^{\pi} = 0^{+}_{1}$, $K^{\pi} =
2^{+}_{1}$ and $K^{\pi} = 0^{-}_{1}$ bands, observed in the
experimental spectrum of $^{24}$Mg.

\section{Theoretical approach}

Consider an $A$-body (one-component) nuclear system, which can be
described in terms of $m=A-1$ translationally invariant relative
Jacobi coordinates $q_{is}$. This allows us to avoid the problem of
the center-of-mass motion from the very beginning. All bilinear
Hermitian combinations of the Jacobi position $q_{is}$ and momentum
$p_{is}$ ($i=1,2,3$; $s=1,2,\ldots, m$) coordinates generate
$Sp(6m,R)$ $-$ the full dynamical group of the whole nuclear system.
The $Sp(6m,R)$ group contains different kinds of possible motions
$-$ collective, internal, cluster, etc., which can be obtained by
reducing it in different ways. By doing this, one performs a
separation of the $3m$ nuclear many-particle fermion variables
$\{q\}$ into kinematical (internal) and dynamical (collective) ones,
i.e. $\{q\} = \{q_{D},q_{K}\}$. According to this, the many-particle
nuclear wave functions can be represented, respectively, as
consisting of collective and internal components \cite{Ganev21c}
\begin{equation}
\Psi(q) = \sum_{\eta} \Theta_{\eta}(q_{D})\chi_{\eta}(q_{K}).
\label{q-WF}
\end{equation}

For example, the (one-component) symplectic symmetry approach to
clustering in atomic nuclei for a two-cluster nuclear system was
firstly introduced by considering the chain \cite{Ganev21c}:
\begin{align}
&Sp(6(A-1),R) \notag\\
\notag\\
&\supset Sp(6,R)_{R} \otimes Sp(6(A_{1}-1),R) \otimes Sp(6(A_{2}-1),R) \notag\\
\notag\\
&\supset Sp(6,R)_{R} \otimes Sp(6,R)_{A_{1}-1} \otimes O(A_{1}-1) \notag\\
\notag\\
&\qquad\qquad\quad \ \otimes Sp(6,R)_{A_{2}-1} \otimes O(A_{2}-1) \notag\\
\notag\\
&\supset \ U_{R}(3) \ \ \otimes \ \ U_{A_{1}-1}(3) \ \otimes \
\ U_{A_{2}-1}(3) \ \ \otimes \ S_{A_{1}} \ \otimes \ S_{A_{2}} \notag\\
&\ \ [n_{R},0,0] \ [n^{C_{1}}_{1},n^{C_{1}}_{2},n^{C_{1}}_{3}] \
[n^{C_{2}}_{1},n^{C_{2}}_{2},n^{C_{2}}_{3}] \quad f_{1} \quad\quad
f_{2} \notag\\
\notag\\
&\supset \ \ U_{R}(3) \ \ \otimes \ \ U_{C}(3) \notag\\
&\quad [n_{R},0,0] \quad [n^{C}_{1},n^{C}_{2},n^{C}_{3}] \notag\\
\notag\\
&\supset \ \ U(3) \ \ \supset \ \ SU(3) \ \ \supset \ \ SO(3) . \label{sp6mR-2cc}\\
&\ [N_{1},N_{2},N_{3}] \ \ (\lambda,\mu) \quad \kappa \quad \ \ L
\notag
\end{align}
The groups $Sp(6,R)_{R}$, $Sp(6(A_{1}-1),R)$ and $Sp(6(A_{2}-1),R)$
are related to the intercluster, first and second cluster system
degrees of freedom, respectively. The intercluster relative motion
of the two clusters is described by one of the $(A-1)$ relative
Jacobi vectors, denote it by $\textbf{q}^{R}$, whereas the rest
$(A-2)$ Jacobi vectors are related to the internal states of the two
clusters. $Sp(6,R)_{A_{\alpha}-1}$ subgroups with $\alpha=1,2$ in
$Sp(6,R)_{A_{\alpha}-1} \otimes O(A_{\alpha}-1) \subset
Sp(6(A_{\alpha}-1),R)$ are the dynamical groups of collective
excitations of the two clusters (referred also to as internal
cluster excitations), while the groups $O(A_{\alpha}-1)$ allow to
ensure the proper permutational symmetries of both clusters.

According to the reduction chain (\ref{sp6mR-2cc}), corresponding to
the case of two-cluster nuclear system ($A=A_{1}+A_{2}$), the
well-known resonating group method (RGM) anzatz \cite{Wildermuth77}
\begin{equation}
\Psi(q) =
\mathcal{A}\{\phi_{1}(A_{1}-1)\phi_{2}(A_{2}-1)f(\textbf{q}^{R})\},
\label{WFRGM}
\end{equation}
can be related to Eq.(\ref{q-WF}) by making the following
identifications: $\chi_{\eta}(q_{K}) = \phi_{1}(A_{1}-1)
\phi_{2}(A_{2}-1)$ and $\Theta_{\eta}(q_{D})=f(\textbf{q}^{R})$. In
the present work, however, we consider the following modified
coupling scheme:
\begin{align}
&Sp(6(A-1),R) \notag\\
\notag\\
&\supset Sp(6,R)_{R} \otimes Sp(6(A_{1}-1),R) \otimes Sp(6(A_{2}-1),R) \notag\\
\notag\\
&\supset Sp(6,R)_{R} \otimes Sp(6,R)_{C} \otimes O(A-2) \notag\\
&\qquad \langle \sigma^{R} \rangle \qquad\quad \ \ \langle
\sigma^{C} \rangle \qquad\qquad  \notag\\
&\qquad \ \cup n_{R}\rho_{R} \qquad \ \ \cup \ n_{C}\rho_{C} \notag\\
&\quad \quad U_{R}(3) \quad \otimes \quad U_{C}(3) \quad \supset \notag\\
&\quad [E^{R}_{1},0,0] \quad [E^{C}_{1},E^{C}_{2},E^{C}_{3}] \notag\\
\notag\\
&\supset \ \ U(3) \qquad\quad \supset \qquad SO(3) , \label{Sp6mR-RSp6xCSp6Am2R}\\
&\ [E_{1},E_{2},E_{3}] \quad \ \kappa \qquad\quad \ L \notag
\end{align}
which turns out to be more convenient for our further considerations
and is directly related to the symplectic symmetry associated with
the cluster effects in atomic nuclei. We note that $Sp(6,R)_{C}
\equiv Sp(6,R)_{C_{1}} \otimes Sp(6,R)_{C_{2}}$ and $O(A-2) \equiv
O(A_{1}-1) \otimes O(A_{2}-1)$ are direct-product groups. If we
consider the $(A-1)$ system as a whole, not separated into clusters,
then the relevant reduction of the full dynamical group
$Sp(6(A-1),R)$ is provided by the chain $Sp(6(A-1),R) \supset
Sp(6,R) \otimes O(A-1)$, which defines the (one-component) $Sp(6,R)$
symplectic collective model \cite{RR1} with six microscopic
collective (dynamical) and $3(A-1)-6$ internal (kinematical) degrees
of freedom \cite{Van74,Filippov81,GC,cdf}. Then the permutational
symmetry is ensured by the symmetric group $S_{A}$ through the
reduction $O(A-1) \supset S_{A}$. Similarly, the permutational
symmetry of the two clusters is ensured by the chains $O(A_{1}-1)
\supset S_{A_{1}}$ and $O(A_{2}-1) \supset S_{A_{2}}$, respectively.

Introducing the standard creation and annihilation operators of
three-dimensional harmonic oscillator quanta
\begin{align}
&b^{\dagger R}_{i}=
\sqrt{\frac{\mu\omega_{R}}{2\hbar}}\Big(q^{R}_{i}
-\frac{i}{\mu\omega_{R}}p^{R}_{i}\Big), \notag\\
&b^{R}_{i}=\sqrt{\frac{\mu\omega_{R}}{2\hbar}}\Big(q^{R}_{i}
+\frac{i}{\mu\omega_{R}}p^{R}_{i}\Big), \label{R-bos}
\end{align}
the group of intercluster excitations $Sp(6,R)_{R}$ can be
represented by means of the following set of generators
\begin{align}
&F^{R}_{ij} = b^{\dagger R}_{i}b^{\dagger R}_{j}, \quad G^{R}_{ij} =
b^{R}_{i}b^{R}_{j}, \label{FGs}\\
&A^{R}_{ij} = \frac{1}{2}\Big(b^{\dagger R}_{i}b^{R}_{j}
+b^{R}_{i}b^{\dagger R}_{j}\Big) \label{As}
\end{align}
in the form $Sp(6,R)_{R}=\{F^{R}_{ij}, G^{R}_{ij},A^{R}_{ij}\}$. In
Eq.(\ref{R-bos}) $\mu=\big(\frac{A_{1}A_{2}}{A_{1}+A_{2}}\big)M$ is
the reduced mass and we chose
$\omega_{R}=\big(\frac{A_{1}+A_{2}}{A_{1}A_{2}}\big)\omega$, so that
for the oscillator length parameter we obtain $b_{0} =
\sqrt{\frac{\hbar}{\mu\omega_{R}}} = \sqrt{\frac{\hbar}{M\omega}}$.
As can be seen, the operators (\ref{FGs}) create or annihilate a
pair of oscillator quanta, whereas the operators (\ref{As}) preserve
the number of quanta and generate the subgroup $U_{R}(3) \subset
Sp(6,R)_{R}$. In this way, the $Sp(6,R)_{R}$ generators of
intercluster excitations can change the number of oscillator quanta
by either 0 or 2. Thus, acting on the ground state by the
$Sp(6,R)_{R}$ generators one can produce only the positive-parity
$SU(3)$ cluster-model states of even oscillator quanta only. The
negative-parity cluster-model states, in turn, consist of odd number
of oscillator quanta and are associated with the $SU(3)$ basis
states of the odd irreps of the group $Sp(6,R)_{R}$. Alternatively,
one may consider the slightly extended semi-direct product group
$[HW(3)_{R}]Sp(6,R)_{R} = \{F^{R}_{ij},
G^{R}_{ij},A^{R}_{ij},b^{\dagger R}_{i},b^{R}_{i},I\}$, consisting
of the $Sp(6,R)_{R}$ and $HW(3)_{R} =
\{b^{\dagger,R}_{i},b^{R}_{i},I\}$ generators. In this way, by
acting on the ground state by the
$WSp(6,R)_{R}=[HW(3)_{R}]Sp(6,R)_{R}$ generators one generates the
cluster-model states of both even and odd number of oscillator
quanta by changing the number of oscillator quanta by 0, 1 and 2.
The latter considerations can be made rigorous by replacing the
group $Sp(6(A-1),R)$ in (\ref{sp6mR-2cc}) by
$[HW(3(A-1))]Sp(6(A-1),R)$, which will be the maximal dynamical
group for the whole $A$-nucleon nuclear system. We note also that
the $WSp(6,R)_{R}=[HW(3)_{R}]Sp(6,R)_{R}$ group contains, in
contrast to the group $Sp(6,R)_{R}$, the $E1$ dipole operator among
its generators, which actually couples the even and odd
$Sp(6,R)_{R}$ irreps.

Similarly one obtains the oscillator realizations for the dynamical
group of collective (internal cluster or major shell) excitations of
the two clusters $Sp(6,R)_{C}$ $= \{F^{C}_{ij}$ $=
\sum_{s=1}^{A-2}b^{\dagger}_{is}b^{\dagger}_{js}$,
$G^{C}_{ij}=\sum_{s=1}^{A-2}b_{is}b_{js}$, $A^{C}_{ij}=
\frac{1}{2}\sum_{s=1}^{A-2}\big(b^{\dagger}_{is}b_{js}
+b_{is}b^{\dagger}_{js}\big)\}$ in terms of the harmonic oscillator
creation and annihilation operators
\begin{align}
&b^{\dagger}_{is}= \sqrt{\frac{M\omega}{2\hbar}}\Big(q_{is}
-\frac{i}{M\omega}p_{is}\Big), \notag\\
&b_{is}=\sqrt{\frac{M\omega}{2\hbar}}\Big(q_{is}
+\frac{i}{M\omega}p_{is}\Big), \label{C-bos}
\end{align}
where $i,j=1,2,3$, $s=1,2,\ldots, A-2$ and $M$ is the nucleon mass.
Actually, this realization of the $Sp(6,R)_{C} \equiv
Sp(6,R)_{C_{1}} \otimes Sp(6,R)_{C_{2}}$ group corresponds to the
case when its $C_{1}X_{C_{1}} + C_{2}X_{C_{2}}$ generators (with
$X_{1} \in Sp(6,R)_{C_{1}}$ and $X_{2} \in Sp(6,R)_{C_{2}}$) are
taken in the form $X_{C_{1}} + X_{C_{2}}$, i.e. when
$C_{1}=C_{2}=1$. This particular case corresponds to the reduction
$Sp(6,R)_{C_{1}} \otimes Sp(6,R)_{C_{2}} \supset
Sp(6,R)_{C_{1}+C_{2}}$. If required, one can consider the symplectic
groups related separately to each cluster, i.e.
$Sp(6,R)_{C_{\alpha}} = \{F^{C_{\alpha}}_{ij}$ $=
\sum_{s=1}^{A_{\alpha}-1}b^{\dagger}_{is}b^{\dagger}_{js}$,
$G^{C_{\alpha}}_{ij}=\sum_{s=1}^{A_{\alpha}-1}b_{is}b_{js}$,
$A^{C_{\alpha}}_{ij}=
\frac{1}{2}\sum_{s=1}^{A_{\alpha}-1}\big(b^{\dagger}_{is}b_{js}
+b_{is}b^{\dagger}_{js}\big)\}$ with $\alpha = 1,2$.

The symplectic basis states of the $R$- or $C$-subsystems are
determined by the $Sp(6,R)_{\alpha}$ ($\alpha=R,C$) lowest-weight
state $|\sigma^{\alpha} \rangle$, defined by
\begin{align}
&G^{\alpha}_{ij} |\sigma^{\alpha} \rangle = 0, \notag\\
&A^{\alpha}_{ij} |\sigma^{\alpha} \rangle = 0, \qquad i<j \notag\\
&A^{\alpha}_{ii} |\sigma^{\alpha} \rangle =
\bigg(\sigma^{\alpha}_{i}+\frac{m_{\alpha}}{2}
\bigg)|\sigma^{\alpha} \rangle, \label{RC-Sp6R-LWS}
\end{align}
and are classified by the following reduction chain
\begin{align}
&Sp(6,R)_{\alpha} \qquad \supset \qquad\qquad U_{\alpha}(3). \label{RC-Sp6R-DC}\\
&\quad \langle \sigma^{\alpha} \rangle \qquad\quad
n_{\alpha}\rho_{\alpha} \qquad
[E^{\alpha}_{1},E^{\alpha}_{2},E^{\alpha}_{3}]_{3} \notag
\end{align}
For the $R$-subsystem $n_{R}=\rho_{R}=m_{R}=1$ and the allowed
$U_{R}(3)$ irreps are only fully symmetric, i.e. of the type
$[E^{R},0,0]_{3}$. Similarly, for the $C$-subsystem $m_{C}=A-2$ and
the $U_{C}(3)$ irreps are of general type
$[E^{C}_{1},E^{C}_{2},E^{C}_{3}]_{3}$. The $Sp(6,R)_{\alpha}$ basis
states can then be represented in the following coupled form
\begin{align}
|\Psi(\sigma^{\alpha}n_{\alpha}\rho_{\alpha}E^{\alpha}\eta_{\alpha})\rangle
= [P^{(n_{\alpha})}(F^{\alpha}) \times |\sigma^{\alpha}
\rangle]^{\rho E^{\alpha}}_{\eta_{\alpha}},
\end{align}
where
$E^{\alpha}=[E^{\alpha}_{1},E^{\alpha}_{2},E^{\alpha}_{3}]_{3}$
denotes the coupled $U_{\alpha}(3)$ irrep and $\rho_{\alpha}$ is a
multiplicity label of its appearance in the product $n \otimes
\sigma^{\alpha}$ with
$n^{\alpha}=[n^{\alpha}_{1},n^{\alpha}_{2},n^{\alpha}_{3}]_{3}$ and
$\sigma^{\alpha}=[\sigma^{\alpha}_{1},\sigma^{\alpha}_{2},\sigma^{\alpha}_{3}]_{3}$.
Finally, the symbol $\eta_{\alpha}$ labels basis states of the group
$U_{\alpha}(3)$. Hence, the total wave functions of the whole
two-cluster system with respect to the whole chain
(\ref{Sp6mR-RSp6xCSp6Am2R}) can be written in an $U(3)$-coupled form
as $\Psi = [\Psi_{R} \times \Psi_{C}]^{\rho E}_{kLM}$ with the
identifications: $\Psi_{R}=f(\textbf{q}^{R})$, $\Psi_{C}= \phi(A-2)$
and $\phi(A-2) = \phi_{1}(A_{1}-1) \phi_{2}(A_{2}-1)$.

Generally, the physical operators of interest (e.g., Hamiltonian and
transition operators) within the fully algebraic approach should be
expressed through the generators of different subgroups in
(\ref{Sp6mR-RSp6xCSp6Am2R}). For instance, the quadrupole generators
of the groups $Sp(6,R)_{R}$ and $Sp(6,R)_{C}$ can be written in the
form
\begin{align}
&Q^{R}_{ij} = q^{R}_{i}q^{R}_{j} = A^{R}_{ij} + \frac{1}{2}\Big(
F^{R}_{ij}+G^{R}_{ij}\Big), \label{QR} \\
&Q^{C}_{ij} = \sum_{s=1}^{A-2}q_{is}q_{js} = A^{C}_{ij} +
\frac{1}{2}\Big(F^{C}_{ij}+G^{C}_{ij}\Big). \label{QC}
\end{align}
Additionally, to obtain the charge quadrupole operators, the cluster
subsystem quadrupole operators $Q^{C}_{ij}$ must be multiplied by
the factor $e_{eff}(Z-1)/(A-2)$. The total charge quadrupole
operator of the whole system, according to the group part
$Sp(6,R)_{R} \otimes Sp(6,R)_{C}$ of (\ref{Sp6mR-RSp6xCSp6Am2R}),
can be represented as
\begin{align}
Q_{ij} = \alpha_{1}Q^{R}_{ij} +\alpha_{2}Q^{C}_{ij}. \label{QRC1C2}
\end{align}
The equal strengths $\alpha_{1}=\alpha_{2}=\alpha$ would correspond
to the chain $Sp(6,R)_{R} \otimes Sp(6,R)_{C} \supset Sp(6,R)$,
which differs from the chain (\ref{Sp6mR-RSp6xCSp6Am2R}) with
obvious identifications $\alpha_{1}=e_{eff}$ and
$\alpha_{2}=e_{eff}(Z-1)/(A-2)$ (i.e. $\alpha_{1} \neq \alpha_{2}$).
Alternatively, according to the group part $U_{R}(3) \otimes
U_{C}(3)$ of (\ref{Sp6mR-RSp6xCSp6Am2R}), one could consider as a
transition operator only the $SU(3)$ components of the charge
quadrupole operator
\begin{align}
\widetilde{Q}_{ij} = e_{eff}A^{R}_{ij} +
e_{eff}\Big(\frac{Z-1}{A-2}\Big)A^{C}_{ij}. \label{SU3QRC1C2}
\end{align}
In spherical coordinates the latter operators become
\begin{align}
\widetilde{Q}_{2m} =\sqrt{3}\bigg(e_{eff}A^{R}_{2m} +
e_{eff}\Big(\frac{Z-1}{A-2}\Big)A^{C}_{2m}\bigg).
\label{SU3QRC1C2-sph}
\end{align}

At this point we want to make a comment, concerning the other
approaches to clustering in atomic nuclei that are more or less
\emph{related} to the symplectic symmetry. The relation of the
$Sp(6,R)$ and $\alpha$-cluster model states has been done in
Refs.\cite{Sp6R-cluster1,Sp6R-cluster2,Sp6R-cluster3}, using
complicated overlap integrals between the symplectic and cluster
bases. In Refs.\cite{Sp6R-cluster1,Sp6R-cluster2} it has been
demonstrated that the $\alpha$-cluster and $Sp(6,R)$ states are
essentially complementary with decreasing overlap with the increase
of the oscillator quanta excitations $2n \hbar\omega$. Sometimes,
subsets of the full set of the $SU(3)$ basis states, contained in
the $Sp(6,R)$ irreducible spaces, are used in the practical
applications. In this respect, the $Sp(2,R) \subset Sp(6,R)$
\cite{Sp2R} and $Sp(4,R) \subset Sp(6,R)$ submodels \cite{Sp4R}
represent the giant resonance collective excitations along the
$z$-direction or the $z$ and $x$ directions, respectively. The
connection of the $\alpha$-cluster and $Sp(2,R)$ states has been
investigated in Refs.\cite{Sp2R-cluster1,Sp2R-cluster2} for the case
of $^{8}$Be, again confirming their complementary nature. The
$Sp(2,R)$ and $Sp(4,R)$ symplectic models of nuclear structure,
however, don't contain an $SO(3)$ subgroup, which requires the usage
of sophisticated angular-momentum projection techniques (often
combined with the complicated generator coordinate method (GCM)
\cite{Griffin57} calculation of the Hamiltonian and norm overlap
kernels). Recently, the no-core symplectic model (NCSpM)
\cite{NCSpM-1,NCSpM-2} with the $Sp(6,R)$ symmetry has been used to
study the many-body dynamics that gives rise to the ground state
rotational band together with phenomena tied to alpha-clustering
substructures in the low-lying states in $^{12}$C
\cite{NCSpM-1,Launey16,Launey21}. Actually, the approach of
Refs.\cite{NCSpM-1,Launey16,Launey21} exploits the symplectic
symmetry, related to clustering, indirectly by using the complicated
technique of Y. Suzuki \cite{Sp6R-cluster2} for computing, just as
in Refs.\cite{Sp6R-cluster1,Sp6R-cluster2,Sp6R-cluster3}, the
overlaps between the complementary symplectic shell model and
cluster model bases to project the cluster wave functions out of the
NCSpM \cite{NCSpM-1,NCSpM-2} (or \emph{ab initio} symmetry-adapted
no-core shell model (SA-NCSM) \cite{Launey16,Dytrych13}) microscopic
wave functions. The NCSpM differs from the standard $Sp(6,R)$
symplectic model of Rosensteel and Rowe \cite{RR1}, proposed for
description of the quadrupole collectivity of atomic nuclei, only by
the type of nuclear interactions exploited in the calculations.
Thus, using the overlap technique \cite{Sp6R-cluster2}, the cluster
wave functions can be projected out of the standard $Sp(6,R)$ model
instead of those of the NCSpM. In this way the symplectic symmetry
of Refs. \cite{NCSpM-1,Launey16,Launey21} is actually related to the
standard $Sp(6,R)$ model \cite{RR1} of quadrupole collectivity and
the cluster effects in atomic nuclei are implicitly included by
considering sufficiently large model symplectic spaces able to
describe spatially expanded nuclear configurations (high-lying np-nh
excitations). In contrast, the microscopic cluster-model type wave
functions within the present SSAC are computed directly, using the
symplectic symmetry (even for the 0p-0h valence subspace). As can be
seen from Eq.(\ref{Sp6mR-RSp6xCSp6Am2R}), the symplectic symmetry
appears at many levels within the present approach and is associated
with both the intercluster and intracluster (internal) nuclear
excitations. We note also that the $SU_{R}(3) \otimes SU_{C}(3)
\supset SU(3)$ underlying substructure of various cluster models,
i.e. the $SU(3)$-based RGM, has first been considered in
\cite{Hecht77}.

The $Sp(6,R)_{R} \otimes Sp(6,R)_{C}$ group structure exploited in a
full account in the present paper, to our knowledge, has never been
used in the literature. In this way, in contrast to the Refs.
\cite{Sp6R-cluster1,Sp6R-cluster2,Sp6R-cluster3,NCSpM-1,Launey16,Launey21}
in which the overlaps between the cluster model and the $Sp(6,R)$
model wave functions have been calculated in the attempt to "unify"
these two models of nuclear excitations (and revealing their
complementary character), the present work introduces a fully
algebraic \emph{cluster model} that is directly \emph{based on} the
symplectic symmetry. Through the $Sp(6,R)_{R} \otimes Sp(6,R)_{C}$
group substructure, the conventional $Sp(6,R)$ (associated now with
the group $Sp(6,R)_{C}$) symplectic excitations, related to the
quadrupole collectivity in atomic nuclei, are also naturally
incorporated in a purely algebraic and self-consistent way within
the present SSAC. In the limiting case of no clustering, the group
of the whole system $Sp(6(A-1),R)$ reduces to $Sp(6,R) \otimes
O(A-1)$, i.e. we obtain the $Sp(6(A-1),R) \supset Sp(6,R) \otimes
O(A-1)$ group structure, by means of which the standard $Sp(6,R)$
symplectic model of quadrupole collectivity is recovered.

\section{Physical operators and basis states}

\subsection{Hamiltonian}

A typical shell-model Hamiltonian within the fully algebraic
symplectic symmetry approach can be represented in the form
\begin{align}
H = H_{0} + V(F,G,A) + H_{res}, \label{H0-V-Hres}
\end{align}
where $H_{0}$ is the three-dimensional harmonic oscillator
Hamiltonian and $V(F,G,A)$ is a potential, which can be expressed by
mean of the symplectic generators $F,G,A$ of the $Sp(6,R)_{R}$ and
$Sp(6,R)_{C}$ groups. In particular, the interaction, related to the
intercluster relative motion, can be represented by means of the
$Sp(6,R)_{R}$ generators only as $V=V(F^{R}_{ij},
G^{R}_{ij},A^{R}_{ij})$. $H_{0}$ corresponds to the mean field and
determines the shell structure of the nuclear system. It orders the
shell-model states with respect to the number of oscillator quanta.
Within a given oscillator shell the oscillator states are
degenerated, but the introduction of the potential $V(F,G,A)$ splits
them in energy. $H_{res}$ is a residual part, not included in
$V(F,G,A)$. We note also that a large class of microscopic
cluster-model Hamiltonians of the type
\begin{align}
H = H_{C_{1}} + H_{C_{2}} + T(q_{R}) + V(q_{C_{1}},q_{C_{2}},q_{R}),
\label{micro-H}
\end{align}
with $q_{C_{1}}=(q_{1}, \ldots, q_{A_{1}-1})$ and $q_{C_{2}}=(q_{1},
\ldots, q_{A_{2}-1})$, can be expressed in the algebraic form
(\ref{H0-V-Hres}) by means of the elements of the corresponding
symplectic dynamical groups.

\subsection{$E2$ transition operator}

As $E2$ transition operators we chose
\begin{align}
T^{E2}_{2M} = \sqrt{\frac{5}{16\pi}}\bigg[e_{eff}Q^{R}_{2M} +
e_{eff}\Big(\frac{Z-1}{A-2}\Big)Q^{C}_{2M} \bigg], \label{TE2-op}
\end{align}
which generally have both (internal) cluster and intercluster
excitation components.
$Q^{R}_{2M}=\sqrt{3}\Big[A^{R}_{2M}+\frac{1}{2}\Big(
F^{R}_{2M}+G^{R}_{2M}\Big)\Big]$ and
$Q^{C}_{2M}=\sqrt{3}\Big[A^{C}_{2M} +\frac{1}{2}\Big( F^{C}_{2M}
+G^{C}_{2M}\Big)\Big]$ are the quadrupole generators of the groups
$Sp(6,R)_{R}$ and $Sp(6,R)_{C}$, respectively. We recall that
$Sp(6,R)_{R}$ represents the intercluster excitations, whereas
$Sp(6,R)_{C}$ is associated with the internal cluster excitations.
In the case of unchanged internal cluster structure, only the
in-shell $SU(3)$ components
$\widetilde{Q}^{C}_{2M}=\sqrt{3}A^{C}_{2M}$ of $Q^{C}_{2M}$, which
does not affect the internal structure, will contribute to the
$B(E2)$ transition probabilities. In present approach, we use the
bare electric charge, i.e. $e_{eff}=e$. We point out that within the
cluster models the full quadrupole operators associated with the
intercluster excitations, i.e. the symplectic $Sp(6,R)_{R}$
generators $Q^{R}_{2M}$ only (without the $Q^{C}_{2M}$ components;
cf. Eq.(\ref{TE2-op})), were used as $E2$ transition operators in
Refs. \cite{Hess18,Hess19} in contrast to the widely exploited
in-shell or $SU(3)$ components \cite{Cseh21,SAQM,SAQM2,SAQM3}. We
note also that in Ref.\cite{Yepez-Martinez12} the in-shell
$SU_{R}(3)$ quadrupole operators $\widetilde{Q}^{R}_{2M}$ are
multiplied by the factor $\Big(\frac{e_{eff}Z}{A}\Big)$, which in
our opinion is redundant since the intercluster motion is associated
with the single relative Jacobi vector $\textbf{q}^{R}$ (hence the
related charge factor of the $R$-subsystem is simply $e_{eff}$ since
$A_{R}=Z_{R}=1$).

\begin{figure}[h!]\centering
\includegraphics[width=70mm]{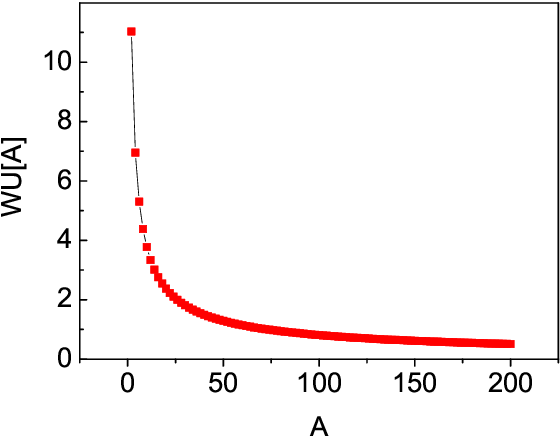}
\caption{(Color online) Function $WU[A]=(1.010
A^{1/6})^{4}\frac{1}{(5.940\times 10^{-2})A^{4/3}}$, representing
the Weisskopf units, as a function of the even values of the mass
number $A$ in the range $2-200$.} \label{WUA}
\end{figure}

We stress that for light nuclei the quadrupole moment $Q_{2M}$
usually is multiplied by the numerical factor
$\sqrt{\frac{5}{16\pi}}$
\cite{Elliott58,Harvey68,Yepez-Martinez12,Mg24a,Mg24b}, i.e. one
uses $T^{E2}_{2M} = \sqrt{\frac{5}{16\pi}}
\Big(\frac{e_{eff}Z}{A}\Big) Q_{2M}$. This factor distinguishes the
form of the quadrupole operator from that which is commonly used in
the shell model, in which $T^{E2}_{2M} =
\Big(\frac{e_{eff}Z}{A}\Big) Q_{2M}$
\cite{de-Shalit63,Brussaard77,Ring80,Heyde94}. We recall that the
units of $Q_{2M}$ are in $b^{2}_{0}$, where
$b_{0}=\sqrt{\frac{\hbar}{M\omega}}$ is the standard oscillator
length parameter. Then, using the bare electric charge $e_{eff}=e$,
one obtains the charge quadrupole moments $\Big(\frac{eZ}{A}\Big)
Q_{2M}$ in units of $eb^{2}_{0}$. Making use of the expression for
the harmonic oscillator length $b_{0}=1.010 A^{1/6}$ $fm$
\cite{Ring80}, the units of charge quadrupole moments become
$efm^{2}$. To obtain the $B(E2)$ transition strengths in Weisskopf
units, one thus must multiply the corresponding expression by the
numerical factor $WU=(1.010 A^{1/6})^{4}\frac{1}{(5.940\times
10^{-2})A^{4/3}}$. This factor as a function of even $A$ values in
the range $2-200$ is shown in Fig. \ref{WUA}, from which it can be
seen that for light nuclei it produces too huge values. In this
respect, the factor $\frac{5}{16\pi}$ entering in the expression for
the $B(E2)$ values will reduce the latter approximately 10 times.
When an effective charge is used, the numerical factor
$\sqrt{\frac{5}{16\pi}}$ is irrelevant. But it is crucial in
obtaining the proper experimentally observed $B(E2)$ values when no
effective charge is used. Otherwise, too huge values are obtained
for the light nuclei due to the large values of the WU factor given
above. For instance, for $A=20-24$, $WU[A] \simeq 2.2$, while for $A
\approx 100$, $WU[A] \simeq 0.8$. This means that the $B(E2)$ values
for light $sd$-shell nuclei with $A=20-24$ will be about three times
larger compared to the $B(E2)$ values for heavy nuclei with $A
\approx 100$. For example, assuming a pure $SU(3)$ structure
$(8,4)$, the $B(E2)$ transition probability produced by the
transition operator (\ref{TE2-op}) without the factor
$\sqrt{\frac{5}{16\pi}}$ is 124.7 W.u., which is too huge compared
to the experimental value $20.3$ W.u. Using a vertical mixing of
different $SU(3)$ states, corresponding to the relative motion
cluster excitations, will increase the $B(E2)$ values even more
dramatically. The same kind of dramatic difference will be obtained
for lighter nuclei from the $s$ and $p$ major shells without the
factor $\sqrt{\frac{5}{16\pi}}$ in the definition of the $E2$
transition operator (\ref{TE2-op}).

\subsection{Basis states}

The basis along the reduction chain (\ref{Sp6mR-RSp6xCSp6Am2R}) can
be written in the form
\begin{align}
|\Gamma;[E_{R},0,0]_{3},[E^{C}_{1}, E^{C}_{2},
E^{C}_{3}]_{3};[E_{1}, E_{2}, E_{3}]_{3};\kappa L \rangle,
\label{basis-0}
\end{align}
where $[E_{R},0,0]_{3}$, $[E^{C}_{1}, E^{C}_{2}$, $E^{C}_{3}]_{3}$,
$[E_{1}, E_{2}, E_{3}]_{3}$, and $L$ denote the irreducible
representations of the $U_{R}(3)$, $U_{C}(3)$, $U(3)$, and $SO(3)$
groups, respectively. The symbol $\Gamma$ labels the set of
remaining quantum numbers of other subgroups in
(\ref{Sp6mR-RSp6xCSp6Am2R}), and $\kappa$ is a multiplicity label in
the reduction $U(3) \supset SO(3)$. Using the standard Elliott's
notations $(\lambda_{R},\mu_{R}) = (E_{R}, 0)$, $(\lambda_{C} =
E^{C}_{1} - E^{C}_{2},\mu_{C} = E^{C}_{2} - E^{C}_{3})$, and
$(\lambda=E_{1}-E_{2},\mu=E_{2}-E_{3})$ for the various $SU(3)$
subgroups and the relation $N=(E_{1}+E_{2}+E_{3})+3(A-1)/2$ for the
number of oscillator quanta (including the zero-point motion), the
basis (\ref{basis-0}) can alternatively be presented as
\begin{align}
|\Gamma N;(E_{R},0),(\lambda_{C},\mu_{C}); (\lambda,\mu);\kappa L
\rangle, \label{cluster-basis}
\end{align}
which turns to be more convenient and will be used in what follows.
The matrix elements of different physical operators then can be
represented in this basis in terms of the $U(3)$ coupling and
recoupling coefficients. The required computational technique for
performing a realistic microscopic calculations is shortly presented
in the Appendix.

\section{Application}

In the present application, we consider only the $^{20}$Ne + $\alpha
\rightarrow$ $^{24}$Mg channel and the ground state configurations
of the two clusters, the latter being unchanged as it is assumed in
the RGM \cite{Wildermuth77}. The $\alpha$ particle represents a
closed-shell nucleus and hence is characterized by the scalar
$SU(3)$ irrep $(0,0)$. The $^{20}$Ne cluster has two protons and two
neutrons in the valence $sd$ shell. Using the supermultiplet
spin-isospin coupling scheme and the codes \cite{UNtoSU3a,UNtoSU3b}
one easily obtains the following Pauli allowed $SU(3)$ states
$(8,0), (4,2), (0,4), (2,0)$. A common practice is to choose the
leading, i.e. maximally deformed $SU(3)$ state, especially for the
strongly deformed nuclei. Hence, for the ground-state cluster
$SU_{C}(3)$ irrep $(\lambda_{C},\mu_{C})$ we obtain
$(\lambda_{C_{1}},\mu_{C_{1}}) \otimes
(\lambda_{C_{2}},\mu_{C_{2}})=(8,0) \otimes (0,0) =(8,0)$. The
Wildermuth condition \cite{Wildermuth77} requires for the minimum
Pauli allowed number of oscillator quanta $E_{R}=8$ of the
intercluster excitations. We note that nonzero values for the
minimal value of $E_{R}$, in contrast to the case of
phenomenological cluster models, in the present microscopic approach
means that we have deformed two-cluster nuclear system that can
rotate, which crucially changes the traditional interpretation of
the $U(3)$ (or, equivalently, $SU(3)$) excitations as being of a
pure vibrational nature. Then the $^{20}$Ne + $\alpha$ model space
is obtained by the outer product $(E_{R},0) \otimes
(\lambda_{C},\mu_{C})$, which produces the following set of $SU(3)$
cluster states:
\begin{align}
(8,0) \otimes (8,0) = &(16,0), (14,1), (12,2), (10,3), \notag\\
&(8,4), (6,5), (4,6), (2,7), (0,8). \label{80x80-product}
\end{align}
To obtain the Pauli allowed cluster states, however, one needs to
compare them to the standard shell-model $SU(3)$ states of the
combined two-cluster system $^{24}$Mg. Using again the
supermultiplet spin-isospin scheme,  the codes
\cite{UNtoSU3a,UNtoSU3b} for 8 valence nucleons in the $sd$ shell
produce the following set: $(8,4), (7,3), (8,1), (4,6), (5,4),
(6,2), (3,5), (4,3), (5,1)$, $(0,8), (2,4), (3,2), (4,0), (1,3),
(0,2)$. The matching $SU(3)$ condition between the latter and the
set (\ref{80x80-product}) defines the Pauli allowed $^{20}$Ne +
$\alpha$ cluster model space for the lowest value $E_{R}=8$: $(8,4),
(4,6), (0,8)$. The relevant $^{20}$Ne + $\alpha$ cluster model space
for the lowest values of $E_{R}$ is given in Table
\ref{Ne20-alpha-IR-space}. The members of each $SU(3)$ multiplet
form a rotational-like sequence of levels, which can be interpreted
as a "cluster band". Similar considerations, for instance, produce
the following Pauli allowed sets of $SU(3)$ states: a) $3(8,4),
3(7,3), (8,1), 2(6,2), (5,1), (4,0)$ and b) $(8,4)$ for the $^{12}$C
+ $^{12}$C $\rightarrow$ $^{24}$Mg and $^{16}$O + $^{8}$Be
$\rightarrow$ $^{24}$Mg channels, respectively.

\begin{table}[h!]
\caption{SU(3) basis states $(\lambda,\mu)$ of the $^{20}$Ne +
$\alpha$ cluster model space for $^{24}$Mg, classified according to
the chain (\ref{Sp6mR-RSp6xCSp6Am2R}), for the lowest values $E_{R}$
of the intercluster excitations. The multiple appearance is denoted
by $\varrho$.} \label{Ne20-alpha-IR-space}
\smallskip\centering\small\addtolength{\tabcolsep}{-0.pt}
\begin{tabular}{l|l|l}
\hline\hline $E_{R}$ & $\hbar \omega $ & $\qquad\qquad
SU(3)~IR^{\prime }s\ \ \varrho (\lambda ,\mu )$
\\ \hline\hline
$\vdots $ & $\vdots $ & $\ldots $ \\ \hline
$10$ & $2$ & $%
\begin{tabular}{l}
$(10,4),(8,5),2(6,6),(9,3),(7,4),(8,2),$ \\
$(4,7),2(2,8),(5,5),(3,6),(4,4),$ \\
$(1,7),(0,6),$%
\end{tabular}%
$ \\ \hline
$9$ & $1$ & $%
\begin{tabular}{l}
$(9,4),(7,5),(8,3),$ \\
$(5,6),(3,7),(4,5),$ \\
$(1,8),(0,7)$%
\end{tabular}%
$ \\ \hline $8$ & $0$ & $(8,4),(4,6),(0,8)$ \\ \hline
\end{tabular}%
\end{table}

In the present application we use the following model Hamiltonian of
algebraic form
\begin{align}
H = H_{DS} + H_{res} + H_{vmix}, \label{HDS-Hres-Hvmix}
\end{align}
where the dynamical symmetry Hamiltonian
\begin{align}
H_{DS} = &H_{0} + \xi C_{2}[Sp(6,R)_{R}] + BC_{2}[SU(3)]  \notag\\
&+ C(C_{2}[SU(3)])^{2} +
\frac{1}{2\mathcal{\mathcal{J}}}C_{2}[SO(3)]  \label{HDS-cluster}
\end{align}
is expressed by means of the Casimir operators of different
subgroups in the chain (\ref{Sp6mR-RSp6xCSp6Am2R}). The $SU(3)$
Casimir operators in the Hamiltonian (\ref{HDS-cluster}) will
arrange the $0\hbar\omega$ $SU(3)$ irreps in energy with the $(8,4)$
multiplet becoming the lowest. In order to account for the
experimentally observed bandhead energies of the lowest $K^{\pi} =
0^{+}_{1}$ and $K^{\pi} = 2^{+}_{1}$ bands contained in the $(8,4)$
multiplet, we introduce also a $K^{2}$ term, i.e. $H_{res} =
bK^{2}$, which is a common practice. In addition, to take into
account the different experimental moments of inertia of various
bands we take them energy dependent, i.e.
$\mathcal{J}=\mathcal{J}_{0}(1+\alpha_{i}E_{i})$, where $E_{i}$ is
the excitation energy of the corresponding bandhead. Such energy-
and/or spin-dependent moments of inertia of the type
$\mathcal{J}=\mathcal{J}_{0}(1+\alpha_{i}E_{i} + \beta L)$ are often
used in the literature, e.g. \cite{SRF,Frauendorf15,Pd102}. Related
to $^{24}$Mg, e.g., energy-dependent moments of inertia were use in
\cite{Mg24c}, whereas in \cite{Mg24b} the third- and fourth-order
$SU(3)$ symmetry-braking interactions $X_{3}$ and $X_{4}$ were used
to split the degeneracy of the $K^{\pi} = 0^{+}_{1}$ and $K^{\pi} =
2^{+}_{1}$ bands, instead of the simpler $K^{2}$ operator (which is
a special linear combination of the operators $X_{3}$, $X_{4}$ and
$L^{2}$) used in the present work and in Ref.\cite{Mg24c}.

\begin{figure}[h!]\centering
\includegraphics[width=70mm]{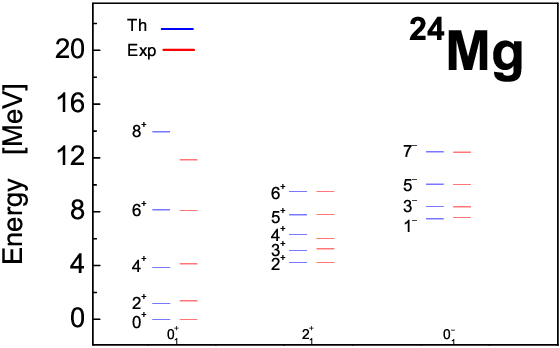}
\caption{(Color online) Comparison of the excitation energies of the
lowest $K^{\pi} = 0^{+}_{1}$, $K^{\pi} = 2^{+}_{1}$ and $K^{\pi} =
0^{-}_{1}$ bands in $^{24}$Mg with experiment \cite{exp}. The values
of the model parameters are: $\xi = 0.314$, $B= -1.203$, $C=
0.0017$, $b = 0.841$, $v_{mix} = -0.0022$ (in MeV), and
$\mathcal{J}_{0}= 2.6$, $\alpha_{2^{+}_{1}} = 0.076$, and
$\alpha_{0^{-}_{1}} = 0.145$ (in MeV$^{-1}$).} \label{Mg24e}
\end{figure}

The second-order $Sp(6,R)_{R}$ Casimir operator splits in energy
different $Sp(6,R)_{R}$ irreducible representations. In particular
it distinguishes between the even and odd $Sp(6,R)_{R}$ irreps
$\langle \sigma^{R} \rangle = \langle \sigma^{R}_{1} +\frac{1}{2},
\sigma^{R}_{2}+ \frac{1}{2}, \sigma^{R}_{3} +\frac{1}{2} \rangle$,
determined by the $Sp(6,R)_{R}$ lowest-weight state labels
$\sigma^{R}\equiv (\sigma^{R}_{1}=E_{R},0,0)$ or
$(\sigma^{R}_{1}=E_{R}+1,0,0)$, which in the present approach
correspond to the positive- and negative-parity cluster-model
states. The lowest even and odd $Sp(6,R)_{R}$ irreps can
conveniently be denoted simply as $0p$-$0h$ $[E_{R}]_{3}$ and
$1p$-$1h$ $[E'_{R}]_{3}$, respectively. They consist respectively of
the following $SU(3)$ shell-model states: $E_{R}, E_{R}+2, E_{R}+4,
\dots$ and $E'_{R}, E'_{R}+2, E'_{R}+4, \dots$ with
$E'_{R}=E_{R}+1$. These two $Sp(6,R)_{R}$ irreps can be considered
as comprising a single irreducible representation of the semi-direct
product group $WSp(6,R)_{R}=[HW(3)_{R}]Sp(6,R)_{R}$ with the set of
$SU(3)$ states: $E_{R}, E_{R}+1, E_{R}+2, E_{R}+3, \dots$. The
$Sp(6,R)_{R}$ Casimir operator within the symplectic representation
$\langle \sigma^{R} \rangle =
\langle\sigma^{R}_{1}+1/2,\sigma^{R}_{2}+1/2,\sigma^{R}_{3}+1/2
\rangle$ takes the following eigenvalue \cite{Van71}:
\begin{equation}
\langle C_{2}[Sp(6,R)_{R}] \rangle = \sum_{i=1}^{3}
\sigma_{i}(\sigma_{i} + 8 - 2i) . \label{Sp6R-EVs}
\end{equation}

The diagonal part of the Hamiltonian (\ref{HDS-Hres-Hvmix}), i.e.
neglecting the mixing term, thus has the following eigenvalues
\begin{align}
E = &N\hbar\omega + \xi\langle
C_{2}[Sp(6,R)_{R}] \rangle + B\langle C_{2}[SU(3)]\rangle \notag\\
& + C(\langle C_{2}[SU(3)]\rangle)^{2}
+\frac{1}{2\mathcal{\mathcal{J}}}L(L+1) + bK^{2}, \label{E-diag}
\end{align}
where $\langle C_{2}[SU(3)]\rangle = 2(\lambda^{2} +\mu^{2}
+\lambda\mu + 3\lambda +3\mu)/3$ is the eigenvalue of the
second-order Casimir operator of $SU(3)$ group.

Finally, we introduce a simple vertical mixing term of the algebraic
form
\begin{align}
H_{vmix} = v_{mix}\Big(A^{R}_{2}\cdot F^{R}_{2} + h.c.\Big),
\label{Hvmix}
\end{align}
which mixes the $SU_{R}(3)$ and, hence, the $SU(3)$ irreducible
representations only vertically. The vertical mixing interaction
(\ref{Hvmix}) corresponds directly to the intercluster excitations.
In this way, we diagonalize the model Hamiltonian
(\ref{HDS-Hres-Hvmix}) in the space of stretched $SU(3)$ cluster
states only, built on the $(8,4)$ multiplet (see Table
\ref{Ne20-alpha-IR-space}), up to energy $20 \hbar\omega$. The
stretched states are the $SU(3)$ cluster states of the type
$(\lambda_{0}+2n,\mu_{0})$ \cite{stretched} with $n = 0, 1, 2,
\ldots$. The results of the diagonalization for the excitation
energies of the lowest $K^{\pi} = 0^{+}_{1}$, $K^{\pi} = 2^{+}_{1}$
and $K^{\pi} = 0^{-}_{1}$ bands in $^{24}$Mg are compared with
experiment \cite{exp} in Fig. \ref{Mg24e}. The values of the model
parameters, obtained by fitting to the excitation energies and the
$B(E2;2^{+}_{1} \rightarrow 0^{+}_{1})$ transition strength, are:
$\xi = 0.314$, $B= -1.203$, $C= 0.0017$, $b = 0.841$, $v_{mix} =
-0.0022$ (in MeV), and $\mathcal{J}_{0}= 2.6$, $\alpha_{2^{+}_{1}} =
0.076$, and $\alpha_{0^{-}_{1}} = 0.145$ (in MeV$^{-1}$). From the
figure one sees a good description of the excitation energies for
the three bands under consideration in $^{24}$Mg. The description of
energy levels of the ground $K^{\pi} = 0^{+}_{1}$ band can be, for
instance, improved further by introducing a spin-dependence of the
ground-state moment of inertia at the price of introducing one more
extra parameter. For instance, in Refs. \cite{Mg24a,Mg24b} an
$L^{4}$-term has been used, which is known to mimics the effect of a
spin-dependent moment of inertia (stretching effect). At first
sight, it may seems that too many parameters are used in the present
microscopic calculations. In this respect, we point out that the two
$K^{\pi} = 0^{+}_{1}$ (ground) and $K^{\pi} = 2^{+}_{1}$ (gamma)
bands are described in Refs. \cite{Mg24b} and \cite{Mg24a} by using
5 and 6 parameters, respectively, which is comparable with the 6
parameters used in the present approach. The extra two parameters in
our calculations are used for the description of the additional
$K^{\pi} = 0^{-}_{1}$ band of negative parity, determining its
experimentally observed excitation bandhead energy and moment of
inertia, respectively.

\begin{figure}[h!]\centering
\includegraphics[width=70mm]{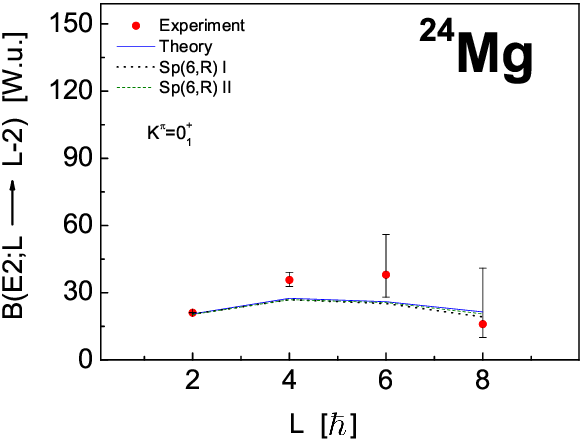}
\caption{(Color online) Comparison of the experimental
\cite{exp,Mg24b,Mg24c} and theoretical intraband $B(E2)$ values in
Weisskopf units between the states of the $K^{\pi} = 0^{+}_{1}$ band
in $^{24}$Mg. For comparison, the theoretical predictions of
Refs.\cite{Mg24a} (Sp(6,R) I) and \cite{Mg24b} (Sp(6,R) II) are
given as well.} \label{be2-Mg24}
\end{figure}

\begin{table}[h!]
\caption{Comparison of the theoretical interband or intraband
$B(E2)$ transition probabilities (in Weisskopf units) for the lowest
states of the $K^{\pi} = 2^{+}_{1}$, and $K^{\pi} = 0^{-}_{1}$ bands
in $^{24}$Mg with the known experimental data \cite{exp,Mg24b,Mg24c}
and the predictions of Refs.\cite{Mg24a,Mg24b} (Sp(6,R) I and
Sp(6,R) II), and those of SACM \cite{Mg24c}. No effective charge is
used in the symplectic based calculations, where for SACM the
adopted values $q_{R}=1.272$ and $q_{C}=1.066$ are used for the
transition operator $T^{E2}_{2M} =
q_{R}\widetilde{Q}^{R}_{2M}+q_{C}\widetilde{Q}^{C}_{2M}$.
$\widetilde{Q}^{R}_{2M}$ and $\widetilde{Q}^{C}_{2M}$ are the
quadrupole operators of the $SU_{R}(3)$ and $SU_{C}(3)$ groups,
respectively.}   
\label{otherBE2s-Mg24}
\smallskip\centering\small\addtolength{\tabcolsep}{2.pt}
\begin{tabular}{lllllll}
\hline\hline $i$ & \multicolumn{1}{|l}{$f$} & \multicolumn{1}{|l}{
$\ \ \ Exp$} & \multicolumn{1}{|l}{$Th$} &
\multicolumn{1}{|l}{$Sp(6,R)\ I$} & \multicolumn{1}{|l}{$Sp(6,R)\
II$} & \multicolumn{1}{|l}{$SACM$} \\ \hline
$2_{2}^{+}$ & $0_{1}^{+}$ & $1.4(0.3)$ & $2.0$ & $\qquad 1.3$ & $\qquad 1.3$ & $\quad 1.6$ \\
$2_{2}^{+}$ & $2_{1}^{+}$ & $2.7(0.4)$ & $4.3$ & $\qquad 1.9$ & $\qquad 2.0$ & $\quad 3.2$ \\
$3_{1}^{+}$ & $2_{1}^{+}$ & $2.1(0.3)$ & $3.6$ & $\qquad 2.3$ & $\qquad 2.2$ & $\quad 2.9$ \\
$4_{2}^{+}$ & $2_{1}^{+}$ & $1.0(0.2)$ & $0.6$ & $\qquad 0.9$ & $\qquad 0.8$ & $\quad 0.6$ \\
$4_{2}^{+}$ & $4_{1}^{+}$ & $1.0(1.0)$ & $5.1$ & $\qquad 2.3$ & $\qquad 2.4$ & $\quad 3.9$ \\
$5_{1}^{+}$ & $4_{1}^{+}$ & $3.9(0.8)$ & $2.3$ & $\qquad 2.0$ & $\qquad 1.9$ & $\quad 2.0$ \\
$6_{2}^{+}$ & $4_{1}^{+}$ & $0.8(_{-0.3}^{+0.8})$ & $0.2$ & $\qquad 1.0$ & $\qquad 0.8$ & $%
\quad 0.3$ \\
$3_{1}^{+}$ & $2_{2}^{+}$ & $34(6)$ & $34.8$ & $\qquad 35.3$ & $\qquad 34.7$ & $\quad 37.5$ \\
$4_{2}^{+}$ & $2_{2}^{+}$ & $16(3)$ & $11.4$ & $\qquad 11.0$ & $\qquad 10.8$ & $\quad 11.4$ \\
$5_{1}^{+}$ & $3_{1}^{+}$ & $28(5)$ & $17.3$ & $\qquad 16.6$ & $\qquad 16.3$ & $\quad 17.5$ \\
$5_{1}^{+}$ & $4_{2}^{+}$ & $14(6)$ & $17.4$ & $\qquad 17.7$ & $\qquad 17.5$ & $\quad 19.5$ \\
$6_{2}^{+}$ & $4_{2}^{+}$ & $23(_{-8}^{+23})$ & $17.6$ & $\qquad 18.3$ & $\qquad 17.7$ & $%
\quad 18.1$ \\
$7_{1}^{+}$ & $5_{1}^{+}$ &  & $18.8$ &  &  & $\quad 19.7$ \\
$8_{2}^{+}$ & $6_{2}^{+}$ & $\geq 3$ & $17.9$ & $\qquad 15.9$ & $\qquad 14.8$ & $\quad 13.7$ \\
$3_{1}^{-}$ & $1_{1}^{-}$ & $<200$ & $23.4$ &  &  & $\quad 32.0$ \\
$5_{1}^{-}$ & $3_{1}^{-}$ & $20(_{-5}^{+8})$ & $25.8$ &  &  & $\quad 34.7$ \\
$7_{1}^{-}$ & $5_{1}^{-}$ & $51(10)$ & $24.3$ &  &  & $\quad 32.3$ \\
\hline
\end{tabular}%
\end{table}

In Fig. \ref{be2-Mg24} we show the intraband  $B(E2)$ transition
probabilities in Weisskopf units between the states of the ground
band, compared with the experimental data \cite{exp} and the
predictions of $Sp(6,R)$ collective model calculations
\cite{Mg24a,Mg24b} (denoted by Sp(6,R) I and Sp(6,R) II,
respectively). From the figure we see practically identical results
for the three symplectic approaches, in which no effective charge is
used (i.e., $e_{eff}=e$).  In addition, in Table
\ref{otherBE2s-Mg24} we compare the known experimental $B(E2)$
values \cite{exp,Mg24b,Mg24c} with the theory for the nonyrast
states of the $K^{\pi} = 2^{+}_{1}$ and $K^{\pi} = 0^{-}_{1}$ bands
in $^{24}$Mg and the predictions of the $Sp(6,R)$ collective model
calculations done in Refs.\cite{Mg24a,Mg24b} and the SACM
calculations of \cite{Mg24c}. We specially compare our present
symplectic based results with those of the $Sp(6,R)$ collective
model. For the quadrupole moment of the excited $2^{+}_{1}$ state we
obtain $Q(2^{+}_{1}) = -0.56$ $eb$, to be compared with the
experimental value $-0.29(3)$ $eb$ \cite{Stone16}, respectively.
From Fig. \ref{be2-Mg24} and Table \ref{otherBE2s-Mg24} one sees
practically equal description of the $B(E2)$ transition strengths
from the states of the $K^{\pi} = 2^{+}_{1}$ and $K^{\pi} =
0^{-}_{1}$ bands in $^{24}$Mg within the different theoretical
approaches. In the present and the $Sp(6,R)$ collective-model
results, no effective charge has been used, in contrast to those of
the SACM approach.

\begin{figure}[h!]\centering
\includegraphics[width=70mm]{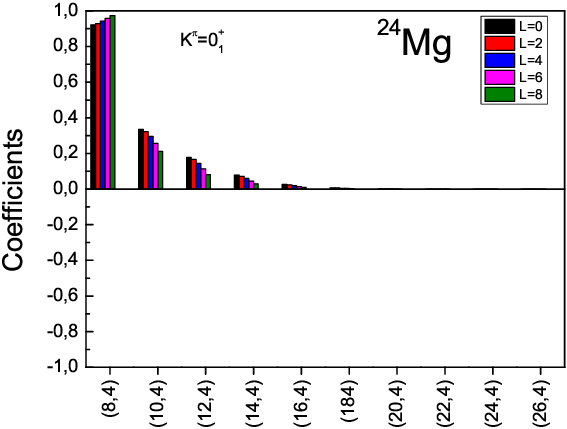}\vspace{5.mm}
\includegraphics[width=70mm]{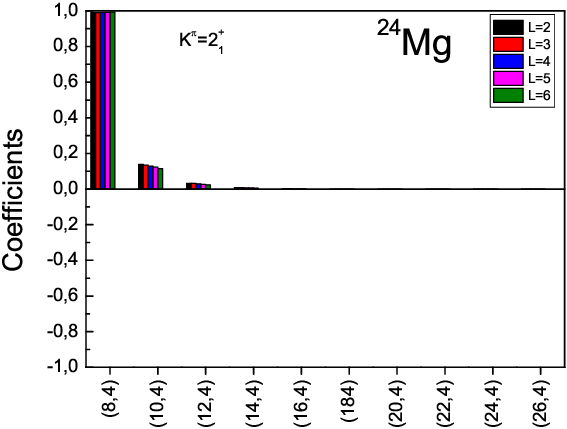}\vspace{5.mm}
\includegraphics[width=70mm]{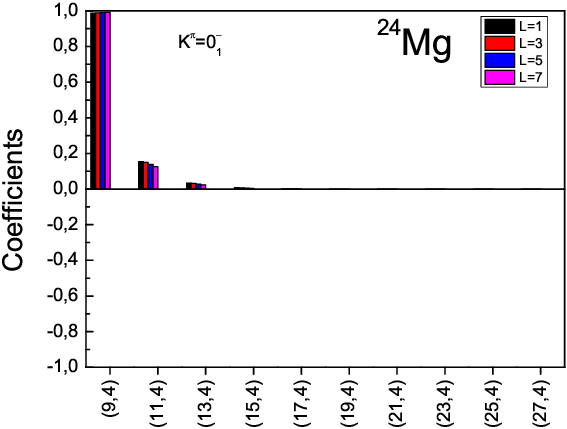}
\caption{$SU(3)$ decomposition of the wave functions for the cluster
states of the lowest $K^{\pi} = 0^{+}_{1}$, $K^{\pi} = 2^{+}_{1}$
and $K^{\pi} = 0^{-}_{1}$ bands in $^{24}$Mg for different angular
momentum values.} \label{SU3dec-Mg24}
\end{figure}

In Fig. \ref{SU3dec-Mg24} we give the $SU(3)$ decomposition of the
wave functions for the cluster-model states of the $K^{\pi} =
0^{+}_{1}$, $K^{\pi} = 2^{+}_{1}$ and $K^{\pi} = 0^{-}_{1}$ bands in
$^{24}$Mg for different angular momentum values. From the figure, we
see a similar structure for the cluster states of the three bands
with predominant contribution of the $0\hbar\omega/1\hbar\omega$
$SU(3)$ multiplet $(8,4)/(9,4)$ and some admixtures due to the
mixing to the excited cluster $SU(3)$ configurations. A similar
microscopic structure for the wave functions for the states of the
$K^{\pi} = 0^{+}_{1}$ and $K^{\pi} = 2^{+}_{1}$ bands is obtained in
Refs.\cite{Mg24a,Mg24b} within the $Sp(6,R)$ collective model. From
the figure, one can see also that the $SU(3)$ decomposition
coefficients are approximately angular-momentum independent
(especially for the states of the $K^{\pi} = 2^{+}_{1}$, and
$K^{\pi} = 0^{-}_{1}$ bands). The latter indicates the presence of a
new type of symmetry, referred to as a quasi-dynamical symmetry in
the sense of Refs.\cite{BR2000,QDSa}. Hence, despite of the mixing
of the $SU(3)$ cluster-model basis states, the microscopic structure
of the experimentally observed cluster states of the three bands
under consideration shows the presence of an approximate $SU(3)$
quasi-dynamical symmetry.

\section{Conclusions}

In the present paper, the recently proposed SSAC has been modified
and applied for the first time to a specific two-cluster nuclear
system, namely to $^{24}$Mg. We have considered the single channel
$^{20}$Ne + $\alpha \rightarrow$ $^{24}$Mg for the microscopic
description of the low-lying cluster states of the lowest $K^{\pi} =
0^{+}_{1}$, $K^{\pi} = 2^{+}_{1}$ and $K^{\pi} = 0^{-}_{1}$ bands in
$^{24}$Mg. According to the RGM \cite{Wildermuth77}, the internal
structure of the two clusters is considered unchanged. This
situation closely resembles the case of different rotational bands
within the framework of different models of nuclear collective
motion, in which the intrinsic structure within the rotational bands
in many cases is considered unchanged. Of course, if needed in the
specific practical applications of the SSAC, one may relax the
requirement of unchanged internal cluster structure as a next
approximation to the nuclear many-body problem.

The purpose of this work was not so to obtain a full and detailed
microscopic description of the low-lying excitation spectrum in
$^{24}$Mg, but to illustrate and test the newly proposed microscopic
SSAC to real nuclear system, consisting of clusters with nonscalar
internal structure that involves more general representations (in
contrast, e.g., to the case of $^{20}$Ne proposed initially in
Ref.\cite{Ganev21c}). This allows to illustrate the more general
nontrivial mathematical structures that appear in the construction
of the microscopic Pauli allowed model space of the $SU(3)$ cluster
states within the present approach.

A good description of the low-lying energy levels for the lowest
$K^{\pi} = 0^{+}_{1}$, $K^{\pi} = 2^{+}_{1}$ and $K^{\pi} =
0^{-}_{1}$ bands in $^{24}$Mg, as well as for the low-energy intra-
and interband $B(E2)$ transition probabilities is obtained within
the SSAC by using a simple dynamical symmetry Hamiltonian to which
residual and vertical mixing terms are also added. This Hamiltonian
is diagonalized within the subspace of stretched $SU(3)$ cluster
states, built on the leading $(8,4)/(9,4)$ multiplet for the
positive/negative-parity states up to energy $20\hbar\omega$. The
microscopic structure of the cluster states of the three bands under
consideration shows a similar picture (cf. Fig.\ref{SU3dec-Mg24})
with a predominant contribution of the $0\hbar\omega$ (for $K^{\pi}
= 0^{+}_{1}$ and $K^{\pi} = 2^{+}_{1}$ bands) and $1\hbar\omega$
(for $K^{\pi} = 0^{-}_{1}$ band) $SU(3)$ irreps, respectively, with
the presence of some admixtures due to the higher cluster $SU(3)$
oscillator configurations. Similarly to the other symplectic based
approaches to nuclear structure, no effective charge is used in the
calculation of the $B(E2)$ transition strengths.

The results obtained in the present paper are shown to be of the
same quality, as those obtained in the microscopic $Sp(6,R)$
collective model calculations performed in Refs.\cite{Mg24a,Mg24b}.
At algebraic level, the present SSAC is to large extent
mathematically equivalent to the $Sp(6,R)$ microscopic collective
model, but physics behind it is different. The latter results in a
different, microscopic cluster interpretation of the observed
low-lying states of the $K^{\pi} = 0^{+}_{1}$, $K^{\pi} = 2^{+}_{1}$
and $K^{\pi} = 0^{-}_{1}$ bands in $^{24}$Mg. Besides the different
interpretation, i.e. the different underlying physics, however, we
want to point one difference between the $Sp(6,R)$ collective model
and the present SSAC that concerns their mathematical structures and
the physical interactions involved in the practical applications of
the two symplectic based approaches. The intrinsic bandhead
structure of the $Sp(6,R)$ collective model in the irreducible
collective subspace of the many-particle Hilbert space contains only
a single $SU(3)$ irrep. This, as we have seen for the case of
$^{24}$Mg, is in contrast to the present SSAC, which for the general
nonscalar internal structure of the clusters naturally leads to a
set of several $SU(3)$ irreducible representations in the
$0\hbar\omega$ subspace. Of course, one can involve different
irreducible $Sp(6,R)$ collective-model spaces but this will require
mixed $Sp(6,R)$ calculations. In order to mix various $Sp(6,R)$
collective-model irreps, one needs to consider components of the
nuclear interaction (e.g., spin-orbit or pairing), which can not be
expressed by the $Sp(6,R)$ generators. In other words, one needs to
involve symplectic symmetry braking nuclear interactions in order to
obtain a set of $SU(3)$ multiplets in the $0\hbar\omega$ subspace.
In contrast, the various $SU(3)$ irreducible representations in the
$0\hbar\omega$ subspace of SSAC can be horizontally mixed by using
an interaction that is expressible, e.g., in terms of the
$Sp(6,R)_{R}$ generators. Additionally, the matrix elements of the
symplectic raising and lowering  (pair creation and annihilation of
oscillator quanta) generators in the SSAC differ from the $Sp(6,R)$
collective-model matrix elements due to the presence of two
symplectic groups, $Sp(6,R)_{R}$ and $Sp(6,R)_{C}$, i.e. the
$Sp(6,R)_{R} \otimes Sp(6,R)_{C}$ subgroup structure, requiring the
$SU(3)$ $9-(\lambda,\mu)$ recoupling coefficients. Due to the
$Sp(6,R)_{R} \otimes Sp(6,R)_{C}$ group, these SSAC matrix elements
in the present formulation differ and from those in the SACM, in
which the $U_{R}(3) \otimes U_{C}(3) \supset U(3)$ subgroup
structure of $Sp(6,R)_{R} \otimes Sp(6,R)_{C}$ is exploited. This
within the SACM leads, however, also to different form of the $E2$
transition operator (an $SU_{R}(3) \otimes SU_{C}(3)$ generator) and
the use of an effective charge. In this respect, using the latter
$SU_{R}(3) \otimes SU_{C}(3)$ generator form for the $E2$ transition
operator, the present SSAC (model space and matrix elements) will
coincide exactly with the SACM in its $SU(3)$ limit only, when
Eq.(\ref{sp6mR-2cc}), i.e. that of Ref. \cite{Ganev21c}, is
exploited.

In short, a new symplectic symmetry approach to clustering in atomic
nuclei is proposed by reducing the dynamical group $Sp(6(A-1),R)$ of
the whole many-nucleon system. The new $Sp(6,R)_{R} \otimes
Sp(6,R)_{C}$ substructure allows to describe simultaneously the
intercluster and intracluster excitations on equal footing by
exploiting the extended $Sp(6,R)$ symplectic symmetry. The $U_{R}(3)
\otimes U_{C}(3) \supset U(3)$ subgroup structure of the
$Sp(6,R)_{R} \otimes Sp(6,R)_{C}$ group further allows to establish
close relationships to the other algebraic cluster models (both
microscopic or phenomenological). Particularly, the relation of the
present SSAC to the SACM has been emphasized here. We note also that
the proper permutational symmetry and the related spin or
spin-isospin content within the present approach, in contrast to the
SACM, for instance, is ensured by the complementary orthogonal group
structure $O(A_{\alpha}-1) \supset S_{A_{\alpha}}$, which together
with the other relevant groups is contained in the $Sp(6(A-1),R)$
dynamical symmetry group of the whole nuclear system. This is
another characteristic feature of the SSAC that distinguishes it
from the SACM, which is known to be a hybrid model of the Elliott's
$SU(3)$ shell model \cite{Elliott58} and the $U(4)$ vibron model of
Iachello \cite{VM}, in which the spin-isospin symmetry is determined
by the complementary Wigner $U_{ST,\alpha}(4)$ ($\alpha=1,2$)
symmetry group \cite{Wigner37}. Finally, the corresponding
computational technique required for performing the practical
applications of the present SSAC to concrete two-cluster nuclear
system is also presented.

The present work opens the path for the further applications of the
SSAC to other two-cluster nuclear systems from different mass
regions. Generally, a horizontal mixing of different $SU(3)$ cluster
states can also be added to the model Hamiltonian, together with the
vertical mixing, in order to perform more sophisticated microscopic
cluster-model calculations within the SSAC. The SSAC can also be
extended to three- and four-cluster nuclear systems. Note that in
the limiting case, the group $Sp(6(A-1),R)$ can be decomposed into
$Sp(6,R)_{1} \otimes Sp(6,R)_{2} \otimes \ldots Sp(6,R)_{A-1}$, in
which the $Sp(6,R)_{s}$ symplectic excitations are associated with
each relative Jacobi vector $\textbf{q}_{s}$ with $s=1, 2, \dots,
A-1$.

\newpage

\appendix*
\section{The SSAC matrix elements}

\onecolumngrid

Generally, if we have a tensor operator $T^{(l_{R},m_{R}) \
L_{R}M_{M}}$ acting on the $R$-subsystem and a tensor operator
$S^{(l_{C},m_{C}) \ L_{C}M_{C}}$ acting on the $C$-subsystem, then
the $SO(3)$-reduced matrix elements of the coupled tensor operator
$[T^{(l_{R},m_{R}) \ L_{R}} \times S^{(l_{C},m_{C}) \
L_{C}}]^{\sigma(l,m) \ L''M''}$ can be expressed in the form
\cite{Millener78,Rosensteel90}:
\begin{align}
&\langle \Gamma; N';(E'_{R},0),(\lambda'_{C},\mu'_{C});
(\lambda',\mu');\kappa' L'||[T^{(l_{R},m_{R}) \ L_{R}} \times
S^{(l_{C},m_{C}) \ L_{C}}]^{(l,m) \ L''M''}||\Gamma;
N;(E_{R},0),(\lambda_{C},\mu_{C}); (\lambda,\mu);\kappa L
\rangle \notag\\
\notag\\
&= \sum_{\rho_{R},\rho_{C},\rho_{f}}\left\{
\begin{tabular}{llll}
$\ (E_{R},0)$ & $(l_{R},m_{R})$ & $\ (E'_{R},0)$ & $\rho_{R}$ \\
$(\lambda_{C},\mu_{C})$ & $(l_{C},m_{C})$ & $(\lambda'_{C},\mu'_{C})$ & $\rho_{C}$ \\
$\ (\lambda,\mu)$ & $\ (l,m)$ & $\ (\lambda',\mu')$ & $\rho_{f}$ \\
$\ \ \ \ 1$ &  $\ \ \ \ \sigma$ &  $\ \ \ \ \ 1 $ &
\end{tabular}%
\right\} \langle (E'_{R},0)|||T^{(l_{R},m_{R}})|||(E_{R},0)
\rangle_{\rho_{R}} \langle
(\lambda'_{C},\mu'_{C})|||S^{(l_{C},m_{C})}|||(\lambda_{C},\mu_{C})
\rangle_{\rho_{C}} \notag\\
\notag\\
&\qquad\qquad\times\langle (\lambda,\mu)\kappa L, (l,m)kL'' ||
(\lambda',\mu')\kappa'L' \rangle_{\rho_{f}}, \label{MEofTS}
\end{align}
where $\{\ldots\}$ and $\langle (\lambda,\mu)\kappa L, (l,m)kL'' ||
(\lambda',\mu')\kappa'L' \rangle$ are the $SU(3)$ recoupling and
coupling coefficients, respectively. In Eq.(\ref{MEofTS}), the
equivalence of the $U(3)$ and $SU(3)$ recoupling coefficients is
used. If the tensor operator acts only in the one subsystem subspace
($R$ or $C$), then the $SU(3)$ $9-(\lambda,\mu)$ recoupling
coefficients reduce to the $6-(\lambda,\mu)$ $U$ coefficients.

As an example consider first the matrix elements with respect to the
whole chain (\ref{Sp6mR-RSp6xCSp6Am2R}) of the tensors $A^{R}_{2M}
\equiv A^{R \qquad\quad 2M}_{(1,1) \ (0,0) \ (1,1)}$ and $F^{R}_{2M}
\equiv F^{R \qquad\quad 2M}_{(2,0) \ (0,0) \ (2,0)}$, entering in
$Q^{R}_{2M}$. For the $SO(3)$-reduced matrix elements of
$A^{R}_{2M}$ we obtain:
\begin{align}
&\langle \Gamma; N;(E_{R},0),(\lambda_{C},\mu_{C});
(\lambda,\mu);\kappa' L'||A^{R}_{2M}||\Gamma;
N;(E_{R},0),(\lambda_{C},\mu_{C}); (\lambda,\mu);\kappa L
\rangle \notag\\
\notag\\
&= \sum_{\rho_{f}}\left\{
\begin{tabular}{llll}
$\ (E_{R},0)$ & $(1,1)$ & $\ (E_{R},0)$ & $1$ \\
$(\lambda_{C},\mu_{C})$ & $(0,0)$ & $(\lambda_{C},\mu_{C})$ & $1$ \\
$\ (\lambda,\mu)$ & $(1,1)$ & $\ (\lambda,\mu)$ & $\rho_{f}$ \\
$\ \ \ \ 1$ &  $\ \ \ 1$ &  $\ \ \ \ 1 $ &
\end{tabular}%
\right\} \langle (E_{R},0)|||A^{R}_{2M}|||(E_{R},0) \rangle \langle
(\lambda,\mu)\kappa L, (1,1)12 ||
(\lambda',\mu')\kappa'L' \rangle_{\rho_{f}} \notag\\
\notag\\
&= \sqrt{\langle2C_{2}[SU_{R}(3)](E_{R},0)\rangle}
\sum_{\rho_{f}}\left\{
\begin{tabular}{llll}
$\ (E_{R},0)$ & $(1,1)$ & $\ (E_{R},0)$ & $1$ \\
$(\lambda_{C},\mu_{C})$ & $(0,0)$ & $(\lambda_{C},\mu_{C})$ & $1$ \\
$\ (\lambda,\mu)$ & $(1,1)$ & $\ (\lambda,\mu)$ & $\rho_{f}$ \\
$\ \ \ \ 1$ &  $\ \ \ 1$ &  $\ \ \ \ 1 $ &
\end{tabular}%
\right\} \langle (\lambda,\mu)\kappa L, (1,1)12 ||
(\lambda',\mu')\kappa'L' \rangle_{\rho_{f}}, \label{MEofA11R}
\end{align}
whereas for the matrix elements of the raising symplectic generators
$F^{R}_{2M}$ we get
\begin{align}
&\langle \Gamma; N+2;(E_{R}+2,0),(\lambda_{C},\mu_{C});
(\lambda',\mu');\kappa' L'||F^{R}_{2M}||\Gamma;
N;(E_{R},0),(\lambda_{C},\mu_{C}); (\lambda,\mu);\kappa L
\rangle \notag\\
\notag\\
&= \left\{
\begin{tabular}{llll}
$\ (E_{R},0)$ & $(2,0)$ & $\ (E_{R}+2,0)$ & $1$ \\
$(\lambda_{C},\mu_{C})$ & $(0,0)$ & $\ \ (\lambda_{C},\mu_{C})$ & $1$ \\
$\ (\lambda,\mu)$ & $(2,0)$ & $\ \ \ (\lambda',\mu')$ & $1$ \\
$\ \ \ \ 1$ &  $\ \ \ 1$ &  $\ \ \ \ \ \ 1 $ &
\end{tabular}%
\right\} \langle (E_{R}+2),0)|||F^{R}_{2M}|||(E_{R},0) \rangle
\langle (\lambda,\mu)\kappa L, (2,0)2 ||
(\lambda',\mu')\kappa'L' \rangle \notag\\
\notag\\
&= \sqrt{(n_{R}+1)(n_{R}+2)}\sqrt{\Delta\Omega(\sigma_{R}
n_{R}'E_{R}';n_{R}E_{R})} \left\{
\begin{tabular}{llll}
$\ (E_{R},0)$ & $(2,0)$ & $\ (E_{R}+2,0)$ & $1$ \\
$(\lambda_{C},\mu_{C})$ & $(0,0)$ & $\ \ (\lambda_{C},\mu_{C})$ & $1$ \\
$\ (\lambda,\mu)$ & $(2,0)$ & $\ \ \ (\lambda',\mu')$ & $1$ \\
$\ \ \ \ 1$ &  $\ \ \ 1$ &  $\ \ \ \ \ \ 1 $ &
\end{tabular}%
\right\} \langle (\lambda,\mu)\kappa L, (2,0)2 ||
(\lambda',\mu')\kappa'L' \rangle, \label{MEofF20R}
\end{align}
where $\sqrt{\Delta\Omega(\sigma_{R} n_{R}'E_{R}';n_{R}E_{R})} =
\sqrt{\Omega(\sigma_{R} n_{R}'E_{R}')-\Omega(\sigma_{R}n_{R}E_{R})}$
and $\Omega(\sigma_{R}n_{R}E_{R}) =\frac{1}{4} \sum_{i=1}^{3}
[2(E^{R}_{i})^{2} -(n^{R}_{i})^{2} + 8(E^{R}_{i}-n^{R}_{i})
-2i(2E^{R}_{i} -n^{R}_{i})]$ \cite{VCS1,VCS2,VCS3}. We note that the
matrix elements of the raising and lowering symplectic generators
are modified, compared to those with respect to the standard
$U_{R}(3) \otimes U_{C}(3) \supset U(3)$ chain, because the
$U_{R}(3)$ and $U_{C}(3)$ groups according to
(\ref{Sp6mR-RSp6xCSp6Am2R}) are embedded in $Sp(6,R)_{R}$ and
$Sp(6,R)_{C}$, respectively. In obtaining the expression
(\ref{MEofF20R}) the standard symplectic computational technique is
used. The $SO(3)$-reduced matrix elements of the lowering symplectic
generators $G^{R}_{2M} \equiv G^{R \qquad\quad 2M}_{(0,2) \ (0,0) \
(0,2)}$ can be obtained from those of the $F^{R}_{2M}$ by conjugate
operation. Similarly, one can obtain the matrix elements of the
quadrupole operators $Q^{C}_{2m}$, acting in the $C$-space.

Finally, we give the $SO(3)$-reduced matrix elements of the tensor
operator $[A^{R}_{2} \times F^{R}_{2}]^{l=0, m=0}_{(2,0) \ (0,0) \
(2,0)}$:
\begin{align}
&\langle \Gamma; N+2;(E_{R}+2,0),(\lambda_{C},\mu_{C});
(\lambda',\mu');\kappa' L||[A^{R}_{2} \times F^{R}_{2}]^{l=0,
m=0}_{(2,0) \ (0,0) \ (2,0)}||\Gamma;
N;(E_{R},0),(\lambda_{C},\mu_{C}); (\lambda,\mu);\kappa L \rangle \notag\\
\notag\\
&= \left\{
\begin{tabular}{llll}
$\ (E_{R},0)$ & $(2,0)$ & $\ (E_{R}+2,0)$ & $1$ \\
$(\lambda_{C},\mu_{C})$ & $(0,0)$ & $\ \ (\lambda_{C},\mu_{C})$ & $1$ \\
$\ (\lambda,\mu)$ & $(2,0)$ & $\ \ \ (\lambda',\mu')$ & $1$ \\
$\ \ \ \ 1$ &  $\ \ \ 1$ &  $\ \ \ \ \ \ 1 $ &
\end{tabular}%
\right\} \langle (E_{R}+2,0)|||[A^{R}_{2} \times F^{R}_{2}]^{l=0,
m=0}_{(2,0) \ (0,0) \ (2,0)}|||(E_{R},0) \rangle \langle
(\lambda,\mu)\kappa L, (2,0)0 || (\lambda',\mu')\kappa'L \rangle,
\notag
\end{align}
where the triple-bared $SU_{R}(3)$ matrix elements can be obtained
by means of the $6-(\lambda,\mu)$ Racah recoupling coefficients:
$\langle (E_{R}+2,0)|||[A^{R}_{2} \times F^{R}_{2}]^{l=0,
m=0}_{(2,0) \ (0,0) \ (2,0)}|||(E_{R},0) \rangle =$ $
U\Big((E_{R},0);(2,0);(E_{R}+2,0);(1,1)\Big|\Big|(E_{R}+2,0);(2,0)\Big)$
$\times\langle (E_{R}+2,0)|||A^{R}_{2M}|||(E_{R}+2,0) \rangle
\langle (E_{R}+2,0)|||F^{R}_{2M}|||(E_{R},0) \rangle$. Using the
latter, one finally obtains for the $SO(3)$-reduced matrix elements
of the tensor operator $[A^{R}_{2} \times F^{R}_{2}]^{l=0,
m=0}_{(2,0) \ (0,0) \ (2,0)}$:
\begin{align}
&\langle \Gamma; N+2;(E_{R}+2,0),(\lambda_{C},\mu_{C});
(\lambda',\mu');\kappa' L||[A^{R}_{2} \times F^{R}_{2}]^{l=0,
m=0}_{(2,0) \ (0,0) \ (2,0)}||\Gamma;
N;(E_{R},0),(\lambda_{C},\mu_{C}); (\lambda,\mu);\kappa L \rangle \notag\\
\notag\\
&= \sqrt{\langle2C_{2}[SU_{R}(3)](n_{R}+2,0)\rangle}
\sqrt{(n_{R}+1)(n_{R}+2)}\sqrt{\Delta\Omega(\sigma_{R}
n_{R}'E_{R}';n_{R}E_{R})} \left\{
\begin{tabular}{llll}
$\ (E_{R},0)$ & $(2,0)$ & $\ (E_{R}+2,0)$ & $1$ \\
$(\lambda_{C},\mu_{C})$ & $(0,0)$ & $\ \ (\lambda_{C},\mu_{C})$ & $1$ \\
$\ (\lambda,\mu)$ & $(2,0)$ & $\ \ \ (\lambda',\mu')$ & $1$ \\
$\ \ \ \ 1$ &  $\ \ \ 1$ &  $\ \ \ \ \ \ 1 $ &
\end{tabular}%
\right\}
\notag\\
&\quad\times
U\Big((E_{R},0);(2,0);(E_{R}+2,0);(1,1)\Big|\Big|(E_{R}+2,0);(2,0)\Big)
\langle(\lambda,\mu)\kappa L, (2,0)0 || (\lambda',\mu')\kappa'L
\rangle. \label{MEofAF20R}
\end{align}
The latter matrix elements allow to perform a numerical
diagonalization of the Hamiltonian (\ref{H0-V-Hres}), containing a
simple vertical mixing term that couples cluster states of different
oscillator quanta. The other matrix elements of interest can be
obtained in a similar way. We recall that computer codes
\cite{code,code2,code3} exist for the numerical calculation of the
required $SU(3)$ coupling and recoupling coefficients.

\twocolumngrid

\end{document}